# Experimental Demonstration of the Sign Reversal of the Order Parameter in (Li$_{1-x}$Fe$_x$)OHFe$_{1-y}$Zn$_y$Se

Zengyi Du[1*], Xiong Yang[1*], Dustin Altenfeld[2*], Qiangqiang Gu[1*], Huan Yang[1], Ilya Eremin[2], Peter J. Hirschfeld[3†], Igor I. Mazin[4], Hai Lin[1], Xiyu Zhu[1], and Hai-Hu Wen[1†]

[1]Center for Superconducting Physics and Materials, National Laboratory of Solid State Microstructures and Department of Physics, Collaborative Innovation Center for Advanced Microstructures, Nanjing University, Nanjing 210093, China

[2] Institut für Theoretische Physik III, Ruhr-Universität Bochum, D-44801 Bochum, Germany

[3] Department of Physics, University of Florida, Gainesville, Florida 32611, USA

[4] Code 6393, Naval Research Laboratory, Washington, DC 20375, USA

[*]These authors contributed equally to this work.

**Iron pnictides are the only known family of unconventional high-temperature superconductors besides cuprates. Until recently, it was widely accepted that superconductivity is spin-fluctuation driven and intimately related to their fermiology, specifically, hole and electron pockets separated by the same wave vector that characterizes the dominant spin fluctuations, and supporting order parameters (OP) of opposite signs[1,2]. This picture was questioned after the discovery**

of a new family, based on the FeSe layers[3-5], either intercalated[6,7] or in the monolayer form[8]. The critical temperatures there reach ~40 K, the same as in optimally doped bulk FeSe – despite the fact that intercalation removes the hole pockets from the Fermi level[9,10] and, seemingly, undermines the basis for the spin-fluctuation theory and the idea of a sign-changing OP. In this paper, using the recently proposed phase-sensitive quasiparticle interference technique, we show that in LiOH intercalated FeSe compound the OP does change sign, albeit within the electronic pockets, and not between the hole and electron ones. This result unifies thepairing mechanism of iron based superconductors with or without the hole Fermi pockets and supports the conclusion that spin fluctuations play the key role in electron pairing.

The discovery of superconductivity in K-intercalated FeSe at a critical temperature on the order of 40 K rekindled interest in Fe-based superconductivity and sent many theorists back to the drawing board[3-5]. Indeed, the simple, transparent and largely accepted idea of spin-fluctuations scattering electron pairs between hole and electron pockets was shaken by the absence of the hole pockets in $K_xFe_{2-y}Se_2$. The fact that the superconducting phase was formed by filamentary inclusions in a strongly magnetic matrix[11] spoke against a conventional single-sign s-wave ("$s^{++}$") pairing[12], and model calculations based on the spin-fluctuation scenario predicted a d-wave state[13,14], which, by symmetry, would have generated gap nodes.[15] On the other hand, later photoemission experiments indicated a nodeless superconducting state[16,17]. After other materials with similar properties were discovered, including $(Li_{1-x}Fe_x)OHFeSe$, which could be synthesized in a single-phase form,[6,7] and FeSe monolayers[8], it became increasingly clear that, with optimal doping, superconductivity at ~40 K and higher is possible without the hole pockets.

We do not discuss here possible mechanisms for this superconductivity, nor even whether it may or may not be similar to the superconductivity in Fe-pnictides. Instead, we will concentrate on a phenomenological question of utmost importance: is superconductivity in FeSe derivatives (assuming they all belong to the same class) of a constant OP sign, or does it have a sign-changing OP? The most natural superconductivity of the latter sort is of $x^2$-$y^2$ type (where x and y are the directions of the Fe-Fe bonds). As discussed in Refs. 5 and 15, crystallographic symmetry lowering due to the Se positions and thus doubling of the unit cell results in this state acquiring gap nodes, although in principle the nodal area may be very small. Moreover, for $K_xFe_{2-y}Se_2$, which exhibits another electron pocket at the zone center, this should also lead to nodal lines on this pocket, now with a typical cosinusoidal angular dependence of the gap. Neither of these effects has been observed[16]. The other type of (truly gapless) sign-changing superconductivity was suggested in Ref. 15 and a detailed theory developed by Khodas and Chubukov[18]. They observed that upon accounting for the spin-orbit interaction, the folded electron ellipses cross and form two pockets that hybridize, of which the inner one is mostly of the xz/yz, and the outer one of xy orbital character. This "bonding-antibonding" scenario[5] postulates that the OP on the inner barrel has one sign, and on the outer one the other. The goal of this paper is not to distinguish between the $d_{x^2-y^2}$ and $s^{\pm}$ symmetries, but to eliminate another popular hypothesis, namely that all electron pockets have the same sign of the OP[14,19]. We emphasize that this question has a principal conceptual importance; it has been widely argued that *no* high temperature superconductivity is possible at all, unless the order parameter averages to (nearly) zero over all Fermi surfaces (thus eliminating the effect of the on-site Coulomb repulsion), and it is generally accepted that spin-fluctuation driven superconductivity necessarily requires a sign-changing order parameter.

Unfortunately, the phase-sensitive probes developed for d-wave cuprates either fail or are more questionable in Fe-based materials. Probes based on Josephson loops with π-contacts, instead of providing a qualitative test, offer only a quantitative probe, since all possible junctions have currents arising from various Fermi surface sheets corresponding to both same sign and opposite sign order parameters[20,21]. Quasiparticle interference (QPI) due to scattering from vortex cores is, in principle, phase-sensitive, but interpretation requires specific models of the superconducting states[22,23]. The technique of identifying with STM bound states formed at a non-magnetic impurity is more promising, and straightforward to measure. However, here the problem is that it is often hard to prove that the investigated impurities are indeed nonmagnetic.

In this paper, we shall report, first, an observation of the above-mentioned bound state, which, notwithstanding the reservations above, strongly suggests a sign-changing order parameter. Second, we will present a set of QPI measurements, and an analysis that does not rely upon the scattering from vortex cores, a poorly understood process, but upon zero-field integrated quantities, as suggested in Ref. 23. As discussed in the next Section, this analysis involves a minimal number of assumptions and utilizes qualitative differences between the integrated QPI intensities for the sign-changing and sign-conserving QP scatterings. This methodology does not require a separation between intra- and inter-pocket scattering, and unambiguously identifies the presence of a sign-changing OP. This universality comes at a cost, though, it cannot distinguish between a nodeless $s^{\pm}$ and a nearly-nodeless d-state in these systems. This choice has to be made based on separate information.

We have measured the scanning tunneling spectra (STS) near a single Fe-substituted impurity in (Li$_{1-x}$Fe$_x$)OHFeSe and show the results in Figure 1. The topographic image measured with $V_{bias}$ = 30 mV, $I_t$ = 100 pA around the single impurity is shown in Fig. 1**a**. One can see that the impurity exhibits a typical image of dumbbell shape with the center located at the Fe site beneath the top Se layer. This dumbbell shape looks quite similar to the non-magnetic Cu impurity in NaFe$_{0.96}$Co$_{0.03}$Cu$_{0.01}$As (Ref. 24), which would hint towards the Zn impurity being the origin of the dumbbell shape. This is also supported by the fact that the simple counting of the impurities visible in the STM agrees well with the analysis of x-ray energy dispersive spectrum (EDX). Both yield about 2 % Zn/Se atomic ratio in the sample, which is, expectedly, less than the nominal 10 % concentration. Indeed, we see essentially one type of surface defect with a concentration in the right range, and, while intrinsic defects like vacancies may or may not occur in a sizeable amount, Zn atoms that exist in the sample must manifest themselves in the STM. Thus, although we cannot definitely identify this impurity as a Zn, it seems exceedingly likely. Displayed in Fig. 1**c** and **1d** are the spatial evolution of the tunneling spectra along the line shown in Fig. **1a** when the magnetic field is zero and 11 T, respectively. One can see that a strong resonant peak appears at about 4.0 meV at the impurity site. When moving away from the impurity, the spectrum recovers a typical shape with two superconducting gaps, similar to the pristine system without Zn (Ref. 25). The two gaps $\Delta_1$ = 14 meV and $\Delta_2$ = 8.5 meV, determined here from spectra far away from the impurity (see the inset of Fig. **1b**) are also quite close to those in the Zn free samples[25]. To illustrate how the magnetic field affects the resonant impurity states, we show in Fig. **1b** two selected spectra measured at the impurity site with and without magnetic field. It is clear that the magnetic field suppresses the resonant peak, but does not shift or split the peak.

Both observations are of great importance; the former proves that the peak in question has superconducting origin, and the latter indicates that the impurity is non-magnetic. Indeed, for a magnetic impurity in the Born limit, the energy of a Bogoliubov-de-Gennes (BdG) quasiparticle contains three terms: $H_{qp} = H_0 + g\mu_B \vec{S} \bullet \vec{H} + \sum_{k,\sigma,k'\sigma'} J\vec{S} \bullet c^+_{k,\sigma}\sigma_{\sigma,\sigma'}c_{k'\sigma'}$ (Ref. 26). Here the first term $H_0$ is just the bare energy of a BdG quasiparticle. The second term arises from the interaction between the magnetic moment **S** and external field **H** and should shift the resonant peak position in magnetic field. The third term is due to the interaction between the BdG quasiparticles and the local magnetic moment, and would have led to a splitting of the peak. We have monitored more than ten individual "dumbbell" impurities and have never observed either a shift or splitting of the peak in present sample, so we conclude the impurity of this type (presumably Zn atom) is non-magnetic. Our conclusion that the bound states are associated with Zn atoms seem at odds with Ref. 27, which found no such states for Zn atoms dosed on the surface. We note, however, that the effective potential for a dosed Zn adatom is significantly weaker than for a substituted Zn impurity. It is highly likely that the former is simply too weak to create a midgap resonance[5,28-30]. This point is discussed further in the Supplementary Information 1.

Earlier angle-resolved photoemission[9,10] and STM measurements[25] have revealed that the hole Fermi surface in (Li$_{1-x}$Fe$_x$)OHFeSe is absent at the Brillouin center ($\Gamma$ point), with only electron pockets at the zone corners. In Figure **2a**, we present a sketch of the Fermi surface topology in the folded (2-Fe) Brillouin zone. Here $\Gamma\widetilde{X}$ is the reciprocal vector of the 2-Fe lattice, that is, it is directed along the Fe-Fe square diagonal and equal to $\sqrt{2}\pi/a$, where a is the Fe-Fe distance.

This Fermi surface geometry generates a rather simple pattern in the QPI image. Since all double-pockets around M-points are equivalent by symmetry (both in the normal and in the superconducting state), there is one roughly circular spot in QPI around the zone center **q**=0, which extends up to $\approx 2\mathbf{k_F}$ (where $\mathbf{k_F}$ is the Fermi vector of the outer pocket). This spot is periodically repeated at each reciprocal lattice vector **G**. Due to tunneling matrix elements, one expects the spot around **G = 0** to be the strongest, the one around **G = (1,0)** weaker (labeled as $\mathbf{q_2}$ in Fig. **2b**), and the one around **G = (1,1)** even weaker ($\mathbf{q_3}$). We will see that this is exactly the picture we observe. The $\mathbf{q_2}$ and $\mathbf{q_3}$ spots do not carry any additional information, being symmetry-equivalent to $\mathbf{q_1}$, so in our analysis we concentrated on the latter. Note that if the spin-orbital coupling induced gap is sufficiently large, one may be able to resolve the $\mathbf{q}_1$ spot into three concentrical rings, corresponding to the threshold of the scattering inside the inner barrel, the outer barrel, and interband scattering. Currently this is beyond attainable resolution, but if in the future it would become possible, in the spirit of the discussion below one will be able to distinguish between the bonding-antibonding $s_\pm$ and quasi-nodeless d.

In Fig. 2**c**, we show a spatial map of the measured differential conductivity $g(\mathbf{r}, E = 8.5 \text{ meV}) = [dI/dV](\mathbf{r}, E = 8.5 \text{ meV})$ around a single impurity, as shown in Fig. 1**a**. One can clearly see patterns generated by the quasiparticle interference (QPI). A Fourier transformation of Fig. 2**c** is shown in Fig. 2**d**, which can be directly compared with the cartoon in Fig. 2**b**. This validates the Fermi surface topology assumed in Fig. 2**a**, although our experimental resolution cannot assess the amplitude of the spin-orbital coupling induced splitting. We emphasize that in order to obtain *dI/dV* on a fine energy mesh, we have used 64×64 grids in the FOV with a single impurity in the center, and measured the STS (-25 mV to 25 mV) at each grid point, masking the remainder of the observation window. Then we can rearrange the $g(\mathbf{r}, E)$ data for all the points to create maps with 64×64 pixels at each energy. This new method ensures required stability in the following phase-related analysis on the QPI intensity. For details of the measurements and spatial position corrections we refer the reader to the Method section and to the Supplementary Information 2.

As discussed, in principle we should have three scattering channels and thus three circles around **q**=0, as marked in Fig. 3**d**. Taking into account the orbital character variation around the Fermi surfaces, we observe that these three rings would roughly correspond to the orbital channels $d_{xy} \leftrightarrow d_{xz/yz}$, $d_{xz/yz} \leftrightarrow d_{xz/yz}$, $d_{xy} \leftrightarrow d_{xy}$ for the interband, inner-intraband and outer-intraband scattering, respectively. Near **q** = 0 we see a bright spot (Fig. 2**b**) that arises from the very short **q**-scattering within one single Fermi surface and between the two Fermi surfaces. Due to the existence of some background QPI intensity in the region close to **q**=0 we shall not consider this part as relevant for the analysis.

Some of us have recently proposed[23] a new methodology for robust determination of possible order parameter sign reversal, using QPI from a single non-magnetic impurity. The central prediction is that the anti-symmetrized intensity of the real part of the FT-QPI is proportional to

$$\delta\rho^-(\mathbf{q}, E) = \text{Re}[\rho(\mathbf{q}, +E) - \rho(\mathbf{q}, -E)] \propto U \cdot \text{Im} \frac{\omega^2 - \Delta_1 \Delta_2}{\sqrt{\omega^2 - \Delta_1^2}\sqrt{\omega^2 - \Delta_2^2}}, \quad (1)$$

for a weak scatterer, and the structure of this result is universal in the sense that it does not depend qualitatively on the strength of the scattering and the details of the electronic structure [23]. Here $\Delta_1$ and $\Delta_2$ are the two gaps associated with the two bands, $U$ is a scalar scattering potential, $\rho(\mathbf{q}, E)$ represents the FT-QPI, which is the FT of the spatial map of the differential conductivity $g(\mathbf{r}, E)$. The prediction for an s$^\pm$ pairing is that this quantity, $\delta\rho^-(\mathbf{q}, E)$, at an energy between the two gaps will be coherently enhanced and hence does not change sign, while for an s$^{++}$ case, this quantity is generally small, with an alternating sign between the two gaps. We have calculated $\delta\rho^-$(**q**, E) from our measured data and show it in Fig. 3**a**. The phase shift has been corrected by taking the impurity as the origin point of the QPI image before the FT, as demonstrated in Ref. 31. The details can be found in Supplementary Information 2. In order to get enhanced signal/noise ratio, we have integrated the data within the ring defined by $0.31\pi/a < |\mathbf{q}| < 0.95\pi/a$, with $a$ the Fe-Fe bond length. The energy dependent behavior of $\delta\rho^-$(**q**, E) is presented in the main panel of Fig. 3**b**. One can see a sharp peak at about 4.0 mV, which is nothing but an impurity resonance (cf. Fig. 1**b)**. This peak is unrelated to the phase relied analysis of QPI and needs to be removed prior to further analysis.

To eliminate this effect, we have masked the central part of the 2D map of d*I*/d*V* with a circle of *R*=3 pixels radius by assuming a parabolic relation $\delta\rho(\omega) = A\omega^2 + B\omega$ . The details about this masking can be found in the Supplementary Information 3. The calculated experimental data of $\delta\rho^-$(**q**, E) after this masking are shown in the main panel of Fig. 3**c**.

We are now ready to compare the experimental data with theoretical predictions of Ref. 23. To begin with, we have simulated the results of the *unmasked* processing presented in Fig. 3b, by integrating the simulated $\delta\rho^-$(**q**, E) over the same ring in the **q**-space. For the parameters we used $E_{imp}$ = 4.0 meV, $\Delta_2$ = 8.5 meV, and $\Delta_1$ = 14.5 meV. The calculated result is shown in the inset of Fig. 3**b** (more details of the calculations are presented in the Method Section and in the Supplementary Information 4). We immediately observe that the s$^\pm$ model reproduces the essential features of the experimental data. One can however, argue that in this case the main difference between the hypotheses is the existence of the impurity resonance in the s$^\pm$ case. To address this objection, we present in Fig. 3c the $\delta\rho^-$(**q**, E) obtained after the removal of the resonance, as discussed above, and, in the inset, the *calculated $\delta\rho^-(\mathbf{q}, E)$* subjected to the same masking procedure. Again, we see a qualitative difference between the two cases, with only the $s^\pm$ calculation reproducing the experimental spectrum (and, in fact reproducing it very well). This gives us strong confidence that gaps change sign between two or more Fermi surface sheets, as shown in Fig. 3**d**. Note that our analysis is not based on a detail, model-dependent comparison between the measurements and simulations, but upon a very qualitative analysis, and the observed agreement hinges exclusively on the fact that the assumed pairing state has a sign-changing order parameter.

We have shown, based on both the observation of an in-gap impurity state for the non-magnetic impurities, and a novel theoretical analysis of quasiparticle interference data, that the order parameter in intercalated FeSe, specifically in $(Li_{1-x}Fe_x)OHFe_{1-y}Zn_ySe$, alternates sign, either between the Fermi surface sheets, or within individual sheets, as illustrated in Fig. 3d. Furthermore, the order parameter sign must affect a considerable fraction of the scattering process, meaning that the two opposite signs are roughly balanced. This puts severe restrictions on the available scenarios. Of those discussed so far, two satisfy this experiment: the bonding-antibonding s-wave state or a nearly-nodeless d-wave state, as discussed, for instance, in Ref. 5. At present, our results do not allow us to distinguish between the two. The latter has an advantage that a clear candidate for pairing glue, namely spin fluctuations resulting from inter-pocket nesting in the unfolded zone (note that such fluctuations were observed, at $q=\{\pi,\pi/2\}$ (Ref.32-34) and calculated, at $q=\{\pi,0.625\pi\}$ (Ref.13) in $K_xFe_{2-y}Se_2$ or other intercalated system, which would have naturally led to a state depicted in Fig. 3**d**, right. A principal problem with this option, however, is that (i) in the related $K_xFe_{2-y}Se_2$ itself it is incompatible with the nodeless Γ-pocket gap[16], so one has to assume different pairing symmetry in these otherwise quite similar materials, and, (ii) the symmetry-required gap nodes at the points where the red and blue Fermi lines intersect in Fig. 3**d**, right, must be very steep, *i.e.*, $d\Delta/dk \gg \Delta/k_F$. The bonding-antibonding $s^{\pm}$ scenario, on the other hand, has an additional advantage in the sense that in this case the sign of the order parameter naturally follows that orbital character of the bands, with the *xy* orbitals carrying one sign, and the *xy/yz* orbitals the other (Fig. 3**d**). This greatly narrows the selection from various possible scenarios in the related systems[35]. Our observation of sign reversal gaps in the only electron Fermi pocket dominated system gives a unification of pairing mechanism, namely the spin fluctuations are the major playing role for electron pairing.

## Methods

I.     **Sample synthesis**

(Li$_{1-x}$Fe$_x$)OHFe$_{1-y}$Zn$_y$Se single crystals were synthesized by the hydrothermal ion-exchange method. First, single crystals with nominal composition K$_{0.8}$(Fe$_{0.9}$Zn$_{0.1}$)$_{2-x}$Se$_x$ were manufactured in advance by the same self-flux method that had been used to synthesize pristine K$_{0.8}$Fe$_{2-x}$Se$_x$ single crystals. Afterwards, 10 ml deionized water, 5 g LiOH (J&K, 99% purity), 0.6 g iron powder (Aladdin Industrial, 99% purity), and 0.3 g selenourea (Alfa Aesar, 99% purity) were added to a 50 mL teflon-linked stainless-steel autoclave. After complete mixing, some pieces of $K_{0.8}(Fe_{0.9}Zn_{0.1})_{2-x}Se_x$ single crystals were added into the mixture as well. Then, the autoclave was sealed and heated up to 120 °C, and the temperature was maintained for 40 to 50 hours. Finally, the $(Li_{1-x}Fe_x)OHFe_{1-y}Zn_ySe$ single crystals were obtained by cooling the autoclave to room temperature. An x-ray energy dispersive spectrum (EDS) analysis using scanning electronic microscope suggests that the composition ratio of Fe:Se is about 1.2:1, and the ratio of Zn:Se is about 2 % as determined from the EDS analysis, indicating a partial substitution of Zn at the Fe sites on the Fe layer.

II.     **STM/STS measurements**.

The STM/STS measurements were carried out in a scanning tunnelling microscope (USM-1300, Unisoku Co., Ltd.) with ultra-high vacuum, low temperature and high magnetic field. The samples were cleaved in an ultra-high vacuum with the base pressure of about $1\times10^{-10}$ torr. Tungsten tips were used during all the STM/STS measurements. To raise the signal to noise ratio, a typical lock-in technique was used with an ac modulation of 0.4 mV and 987.5 Hz.

III.    **Differential conductivity measurements on a dense energy mesh in real space.**

In order to eliminate errors introduced during the QPI analysis, we have measured the full spectrum for a dense energy mesh at each point in the real space. First, we choose an area of 6 nm × 6 nm with a single impurity siting at the center of the image. Then, we divide the scanning area into 64 × 64 pixels and measure tunneling spectra in the voltage window from -25 mV to 25 mV with the same set-point at each position. Then the 2D mapping images of the differential conductivity at different energies in the real space are obtained.

IV. Theoretical calculation

To describe the FT-QPI result we employ a two band tight binding parametrization of the two elliptic electron pockets of the Fermi surface, rotated with respect to each other by 90° on the Fermi surface, as described in the Supplementary Information 4. The scattering of quasiparticles by a non-magnetic impurity, measured in the FT-QPI, was calculated as a correction to the local density of states (LDOS), using the standard T-matrix approach describing multiple scattering by a single impurity. In particular, we compute the anti-symmetrized correction to the local density of states

$$\delta\rho^-(\omega) = \frac{1}{2}\text{Tr Im} \sum_{\mathbf{k},\mathbf{q},\mu,\nu} \tau_3 G^0_\mu(k,\omega)\, t_{\mu\nu}(\omega) G^0_\nu(k,\omega) \tag{1}$$

Here, $G^0_\mu(k,\omega) = -\frac{i\omega\tau_0 + \varepsilon_\mu(\mathbf{k})\tau_3 + \Delta_\mu\tau_1}{\omega^2 + \Delta_\mu^2 + \varepsilon_\mu^2(\mathbf{k})}$ refers to the Nambu-Gor'kov Green's function for the band $\mu$, $\tau_i$ is the $i$-th Pauli matrix. $\varepsilon_\mu(\mathbf{k})$ and $\Delta_\mu$ refer to the quasiparticle energy and superconducting gap, respectively, of the corresponding band. The T-matrix for the multiple scattering by a single impurity in the band and Nambu-Gor'kov space is defined as

$$\hat{t}(\omega) = \left[1 - \hat{U}\tau_3 \sum_\mathbf{k} \hat{G}(\mathbf{k},\omega)\right]^{-1} \tau_3 \hat{U}, \tag{2}$$

where $U_{\mu\mu} = U_{intra}\tau_0$ and $U_{\mu\nu} = U_{inter}\tau_0$ are the intra- and interband impurity scattering strengths, respectively.

**Acknowledgements**

We acknowledge the useful discussions with Gabriel Kotliar, Piers Coleman, Dung-Hai Lee, Jun Zhao. The work in NJU was supported by the Ministry of Science and Technology of China (grant number: 2016YFA0300400), national natural science foundation of China (NSFC) with the projects: A0402/11534005, A0402/11190023, A0402/11374143. P.J.H. was supported by NSF-DMR-1407502. I.I.M. was supported by ONR through the NRL basic research program. D.A. and I.E. were supported by the joint DFG-ANR Project (ER 463/8-1) and DAAD PPP USA N57316180.

**Author contributions**

The low-temperature STS measurements were performed by Z.D, X.Y, Q.G, H.Y.. Data analysis was done by Z.D., X.Y., Q.Q.G., H.Y., H.H.W. The samples were grown by H.L. and X.Y.Z. The theoretical calculations were done by D.A. and I.E. All authors contributed to the writing of the paper, with P.H., I.I.M., H.-H.W. responsible for the final text. H.-H.W. coordinated the whole work. All authors have discussed the results and the interpretations.

**Competing financial interests**

The authors declare that they have no competing financial interests.

*Correspondence and requests for materials should be addressed to

hhwen@nju.edu.cn, pjh@phys.ufl.edu

**References**

1. Mazin, I. I., Singh, D. J., Johannes, M. D., & Du, M. H. Unconventional superconductivity with a sign reversal in the order parameter of LaFeAsO$_{1-x}$F$_x$. *Phys. Rev. Lett.* **101**, 057003 (2008).

2. Kuroki, K. *et al.* Unconventional pairing originating from the disconnected Fermi surfaces of superconducting LaFeAsO$_{1-x}$F$_x$. *Phys. Rev. Lett.* **101**, 087004 (2008).

3. Guo, J. G. *et al*. Superconductivity in the iron selenide K$_x$Fe$_2$Se$_2$ (0<x<1.0). *Phys. Rev. B* **82**, 180520 (2010).

4. Fang, M. H. *et al*. Fe-based superconductivity with T$_c$=31K bordering an antiferromagnetic insulator in (Tl,K)Fe$_x$Se$_2$. *Europhys. Lett.* **94**, 27009 (2011).

5. Hirschfeld, P. J., Korshunov, M. M., Mazin, I. I. Gap symmetry and structure of Fe-based superconductors. *Rep. Prog. Phys.* **74**, 124508 (2011).

6. Lu, X. F. *et al*. Coexistence of superconductivity and antiferromagnetism in (Li$_{0.8}$Fe$_{0.2}$)OHFeSe. *Nature Mater*. **14**, 325-329 (2015).

7. Pachmayr, U. *et al*. Coexistence of 3d-Ferromagnetism and Superconductivity in [(Li$_{1-x}$Fe$_x$)OH](Fe$_{1-y}$Li$_y$)Se. *Angew. Chem. Int. Ed*. **54**, 293-297 (2015).

8. Wang, Q. Y. *et al.* Interface-induced high-temperature superconductivity in single unit-cell FeSe films on SrTiO$_3$. *Chin. Phys. Lett.* **29**, 037402 (2012).

9. Niu, X. H. *et al*. Surface electronic structure and isotropic superconducting gap in (Li$_{0.8}$Fe$_{0.2}$)OHFeSe. *Phys. Rev. B* **92**, 060504 (2015).

10. Zhao, L. *et al*. Common electronic origin of superconductivity in (Li, Fe)OHFeSe bulk superconductor and single-layer FeSe/SrTiO$_3$ films. *Nature Commun*. **7**, 10608 (2016).

11. Ding, X. X. *et al.* Influence of microstructure on superconductivity in K$_x$Fe$_{2-y}$Se$_2$ and evidence for a new parent phase K$_2$Fe$_7$Se$_8$. *Nature Comm.* **4**, 1897 (2013).

12. Mazin, I. I. Iron superconductivity weathers another storm, Physics **4**, 26 (2011).

13. Maier, T. A., Graser, S., Hirschfeld, P. J. and Scalapino, D. J. d-wave pairing from spin fluctuations in the KFe$_2$Se$_2$ superconductors. *Phys. Rev. B* **83**, 100515 (2011).

14. Wang, F. *et al*. The Electron Pairing of K$_x$Fe$_{2-y}$Se$_2$. *Europhys. Lett.* **93**, 57003 (2011).

15. Mazin, I. I. Symmetry analysis of possible superconducting states in K$_x$Fe$_y$Se$_2$ superconductors. *Phys. Rev. B* **84**, 024529 (2011).

16. Mu, X. *et al.* Evidence for an s-wave superconducting gap in K$_x$Fe$_{2-y}$Se$_2$ from angle-resolved photoemission. *Phys. Rev. B* **85**, 220504 (2012).

17. Zhao, L. *et al*. Common Fermi-surface topology and nodeless superconducting gap of K$_{0.68}$Fe$_{1.79}$Se$_2$ and (Tl$_{0.45}$K$_{0.34}$)Fe$_{1.84}$Se$_2$ superconductors revealed via angle-resolved photoemission. *Phys. Rev. B* **83**, 140508 (2011).

18. Khodas, M., Chubukov, A. V. Interpocket Pairing and Gap Symmetry in Fe-Based Superconductors with Only Electron Pockets. *Phys. Rev. Lett.* **108**, 247003 (2012).

19. Onari, S., and Kontani, H. Self-Consistent Vertex Correction Analysis for Iron-Based Superconductors: Mechanism of Coulomb-Interaction-Driven Orbital Fluctuations. *Phys. Rev. Lett*. **109**, 137001 (2012).

20. Golubov, A. A., Mazin, I. I. Designing phase-sensitive tests for Fe-based superconductors. *Appl. Phys. Lett*. **102**, 032601 (2013).

21. Parker, D., and Mazin, I. I. Possible Phase-Sensitive Tests of Pairing Symmetry in Pnictide Superconductors. *Phys. Rev. Lett.* **102**, 227007 (2009).

22. Hanaguri, T., Niitaka, S., Kuroki, K., Takagi, H. Unconventional s-Wave Superconductivity in Fe(Se,Te). *Science* **328**, 474-476 (2010).

23. Hirschfeld, P. J., Altenfeld, D., Eremin, I. & Mazin, I. I. Robust determination of superconducting gap sign changes via quasiparticle interference. *Phys. Rev. B* **92**, 184513 (2015).

24. Yang, H. *et al*. In-gap quasiparticle excitations induced by non-magnetic Cu impurities in Na(Fe$_{0.96}$Co$_{0.03}$Cu$_{0.01}$)As revealed by scanning tunnelling spectroscopy. *Nature Commun*. **4**, 2749 (2013).

25. Du, Z. Y. *et al*. Scrutinizing the double superconducting gaps and strong coupling pairing in (Li$_{1-x}$Fe$_x$)OHFeSe. *Nature Commun*. **7**, 10565 (2016).

26. Balatsky, A. V., Vekhter, I. & Zhu, J.-X. Impurity-induced states in conventional and unconventional superconductors. *Rev. Mod. Phys.* **78**, 373-433 (2006).

27. Yan, Y. J. *et al*. Surface electronic structure and evidence of plain s-wave superconductivity in (Li$_{0.8}$Fe$_{0.2}$)OHFeSe. *Phys. Rev. B* **94**, 134502 (2016).

28. Kariyado, T. & Ogata, M. Single Impurity Problem in Iron-Pnictide Superconductors. *J. Phys. Soc. Jpn.* **79**, 083704 (2010).

29. Bang, Y., Choi, H.-Y., Won, H. Impurity effects on the $\pm$s-gap state of the Fe-based superconductors. *Phys. Rev. B* **79**, 054529 (2009).

30. Beaird, R., Vekhter, I., Zhu, J. X. Impurity states in multiband s-wave superconductors: Analysis of iron pnictides. *Phys. Rev. B* **86**, 140507 (2012).

31. Sprau, P. O. *et al*. Discovery of Orbital-Selective Cooper Pairing in FeSe. Preprint at https://arxiv.org/abs/1611.02134 (2016).

32. Davies, N. R. *et al*. Spin resonance in the superconducting state of $Li_{1-x}Fe_xODFe_{1-y}Se$ observed by neutron spectroscopy. *Phys. Rev. B* **94**, 144503 (2016).

33. Pan, B. Y. *et al*. Structure of spin excitations in heavily electron-doped $Li_{0.8}Fe_{0.2}ODFeSe$ superconductors. Preprint at https://arxiv.org/abs/1608.01204 (2016).

34. Park, J. T. *et al.* Magnetic Resonant Mode in the Low-Energy Spin-Excitation Spectrum of Superconducting $Rb_2Fe_4Se_5$ Single Crystals. *Phys. Rev. Lett.* **107**, 177005 (2011).

35. Huang, D., Hoffman, J. E. Monolayer FeSe on $SrTiO_3$. Preprint at https://arXiv.org/abs/1703.09306 (2017).

**Figures and captions**

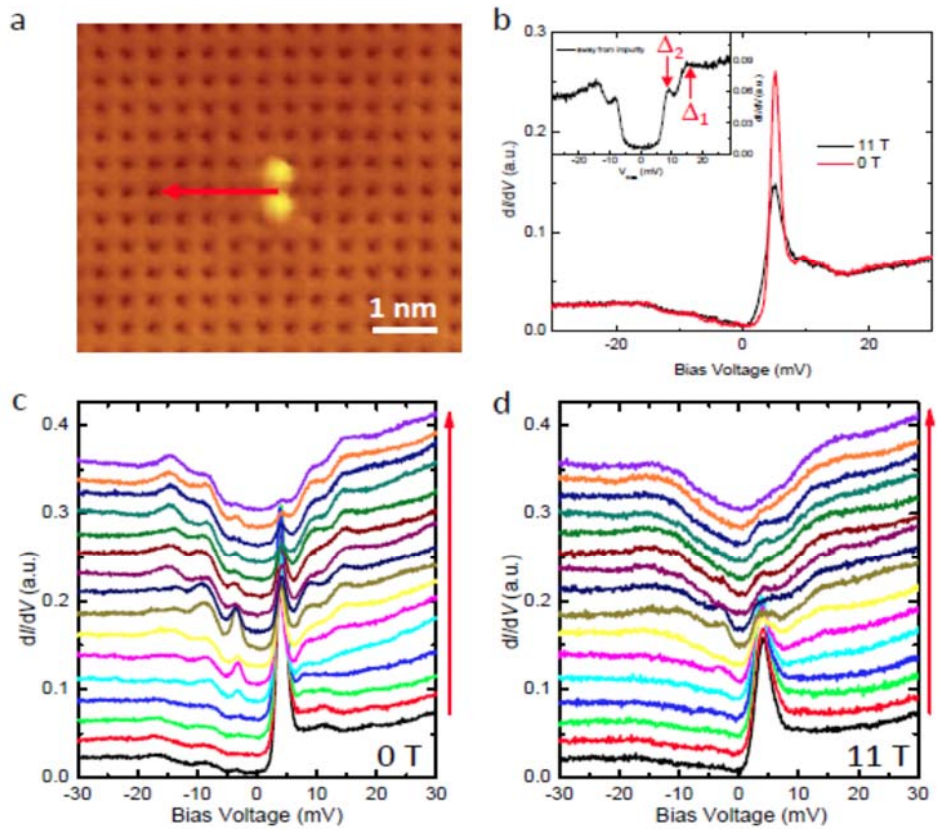

**Figure 1 | In-gap resonant states of a non-magnetic impurity. a,** STM topography ($V_{bias}$ = 30 mV, $I_t$ = 100 pA) around a single impurity. **b,** Tunneling spectra measured at the center of the defect with a dumbbell shape under **B** = 0 T and 11 T, respectively. The peak position of the impurity bound state does not shift under an 11 T magnetic field, indicating its non-magnetic origin. The inset shows a spectrum measured far away from the impurity site with the two gaps $\Delta_1$ = 14 meV, $\Delta_2$ = 8.5 meV, as marked by the arrows. **c, d**, Tunneling spectra measured along the red line in **a** at **B** = 0 T and 11 T, respectively.

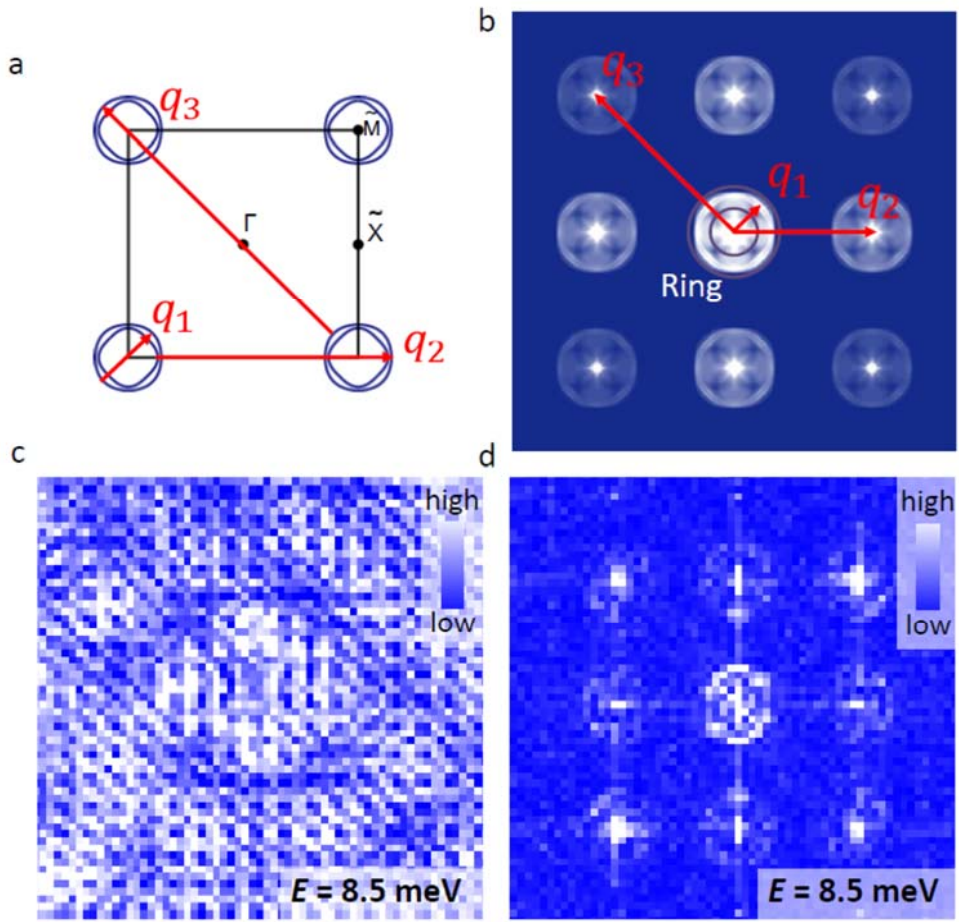

**Figure 2| Quasiparticle Interference around the single impurity. a,** The schematic Fermi surface of $(Li_{1-x}Fe_x)OHFe_{1-y}Zn_ySe$ and three scattering processes in *k* space. The eccentricity of the ellipses is about 0.55. The first Brillouin zone is shown by the solid black line, *i.e.,* there are two overlapping pockets per zone. **b,** A sketch showing the QPI structure expected for this Fermi surface; note that **q₂** and **q₃** differ from **q₁** by the first and second reciprocal lattice vector, respectively. While within the theory employed in this paper they are all equivalent, in reality we expect the signal for larger reciprocal lattice vectors to be weaker, as reflected in this sketch. The sketch is drawn by the self-correlation using the Fermi surface in **a. c,** A typical d*I*/d*V* map $g(\mathbf{r}, E = 8.5 \text{ meV})$ measured at the smaller superconducting gap energy ($\Delta_2$ = 8.5 meV ) around a single impurity. **d,** FT image of **c**, which is qualitatively similar to the sketch shown in **b**.

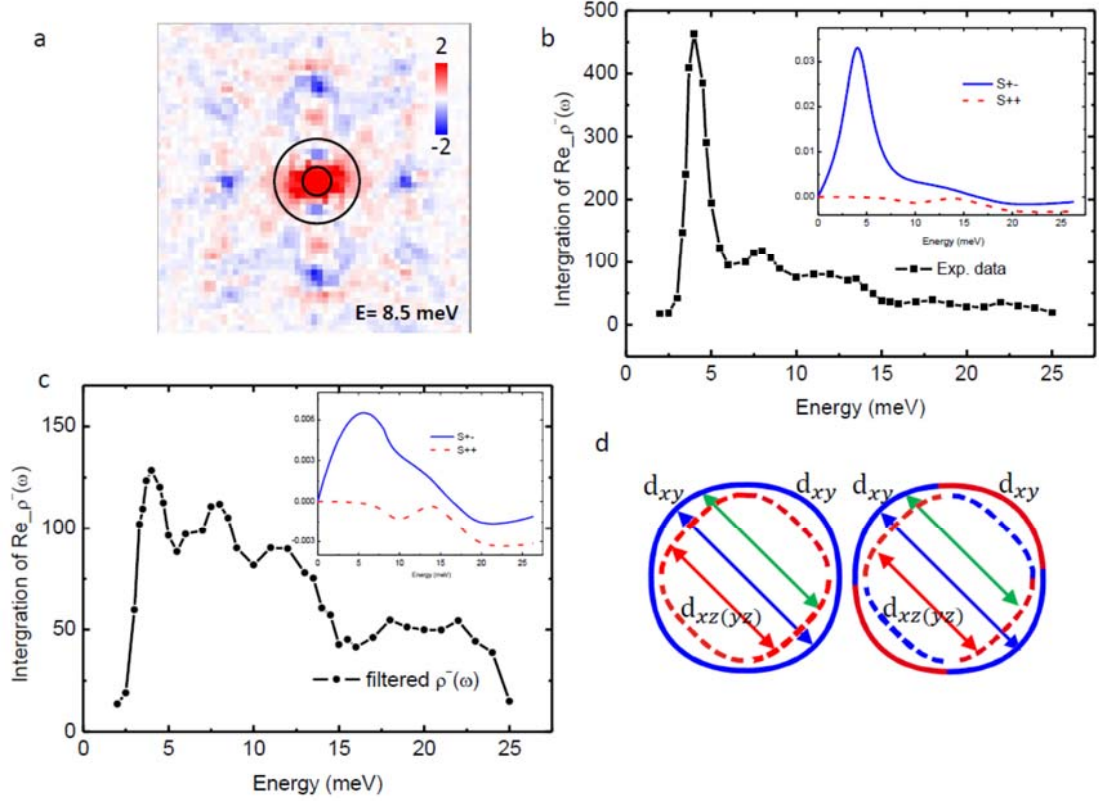

**Figure 3| Determination of sign reversal of the superconducting order parameter. a,** The real-part difference of the FT-QPI $\delta\rho^-(\mathbf{q}, E = 8.5\text{meV}) = \text{Re}[\rho(\mathbf{q}, E = +8.5\text{meV}) - \rho(\mathbf{q}, E = -8.5\text{meV})]$. **b,** Integral of $\delta\rho^-(\mathbf{q}, E)$ versus $E$. The integral range of $|\mathbf{q}|$ is illustrated by the ring in **a** with specified $|\mathbf{q}|$ range $0.31\pi/a < |\mathbf{q}| < 0.95\pi/a$, with $a$ the Fe-Fe bond length. The strong peak at about 4 meV is due to an impurity bound state. The calculated $\delta\rho^-(\mathbf{q}, E)$ is shown in the inset by a blue solid line for s$^\pm$ and red dashed line for s$^{++}$. **c,** Integral of $\delta\rho^-(\mathbf{q}, E)$ as a function of $E$ with the bound state peak masked out as discussed in the main text. The theoretical results are again presented as an inset, with the same convention. **d,** Sketch of the Fermi surface electronic pockets, illustrating two possible sign-reversal scenarios; the blue and red colors represent the opposite signs of the order parameter. The red and blue scattering vectors span same-sign order parameters, and the green ones the opposite sign ones, in both cases.

# SUPPLEMENTARY INFORMATION

# Experimental Demonstration of the Sign Reversal of the Order Parameter in $(Li_{1-x}Fe_x)OHFe_{1-y}Zn_ySe$

Zengyi Du[1*], Xiong Yang[1*], Dustin Altenfeld[2*], Qiangqiang Gu[1*], Huan Yang[1], Ilya Eremin[2], Peter J. Hirschfeld[3†], Igor I. Mazin[4], Hai Lin[1], Xiyu Zhu[1], and Hai-Hu Wen[1†]

**1. Magnetic impurities in $(Li_{1-x}Fe_x)OHFeSe$.**

To contrast our results to those in the samples without Zn-doping, we show here resonant states induced by magnetic impurities in $(Li_{1-x}Fe_x)OHFeSe$. About half of all impurities in $(Li_{1-x}Fe_x)OHFeSe$ behave as the non-magnetic ones, i.e., the resonant peak position seems independent of magnetic field up to 11 T. However, the situation is different for the other half. An example of such magnetic impurities is shown in the Fig. S1. This impurity is also residing at an Fe site in the FeSe layer. Now, two strong separated resonance peaks are observed, located at +3.0 and +6.1 mV, respectively. After a magnetic field of 11 T was applied, the peaks move to +4.5 and +7.3 mV, representing shifts of 1.5 and 1.2 mV. The calculated magnetic moment for this impurity is $M=0.96$ $\mu_B$, and $\Delta E=MH$. It should be noted that the resonance peak at about -6 mV also moves away from zero with magnetic field, which is similar to the situation with Fe-vacancies in $K_xFe_{2-y}Se_2$ [S1]. Although almost half of the impurities in pristine $(Li_{1-x}Fe_x)OHFeSe$ single crystals behave as magnetic, we have not found any such impurities in $(Li_{1-x}Fe_x)OHFe_{1-y}Zn_ySe$ samples. This strongly suggests that the impurities investigated in present work are non-magnetic.

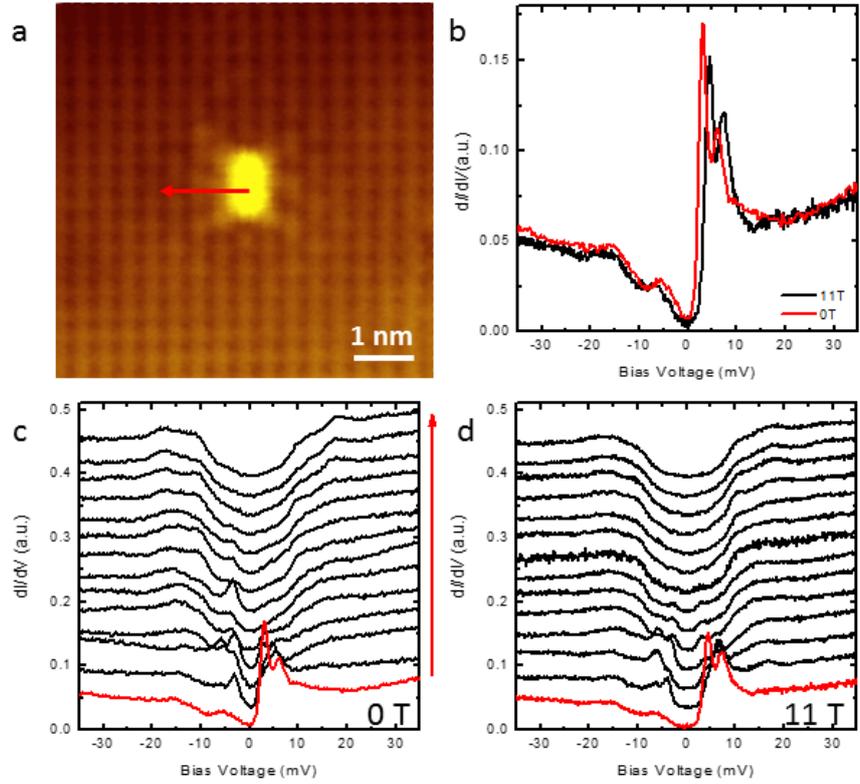

**Figure S1 | In-gap resonant states induced by a magnetic impurity in (Li$_{1-x}$Fe$_x$)OHFeSe.**

**a,** STM topography ($V_{bias}$ = 30 mV, $I_t$ = 100 pA) of a single impurity in (Li$_{1-x}$Fe$_x$)OHFeSe without Zn-doping. **b,** Tunneling spectra measured at the center of the dumbbell-shape impurity under **B** = 0 T and 11 T respectively. **c, d** Spectra measured along the red line in **a** at **B** = 0 T and 11 T.

## 2. Extracting the real-part of the anti-symmetrized FT-QPI spectra

If an impurity is located away from the origin, an additional phase term in **q**-space appears, obscuring the behavior of the Fourier transform: $\mathrm{FT}[f(\mathbf{r} - \mathbf{r_0})] = e^{-i\mathbf{q}\cdot\mathbf{r_0}}\mathrm{FT}[f(\mathbf{r})]$. To avoid this situation, a phase correction is required. This knowledge is well notified in ref. (S2). The technique used is as follows. We first center our image roughly at the impurity site, and take this point as the origin, then measure the spatial conductance map with 64×64 pixels for each energy (examples are shown in Fig. S2**a** and **c**). In order to put the impurity site as the real origin, we remove some edge pixels, and end up with an array of 60×60 pixels, as shown in Fig. S2**b** and **d**. This is then taken as the final image and subject to the Fourier transform. The result is anti-symmetrized and we obtain the tunneling conductance $\delta\rho^-(\mathbf{q},\omega)$ shown in Fig. S2**f**.

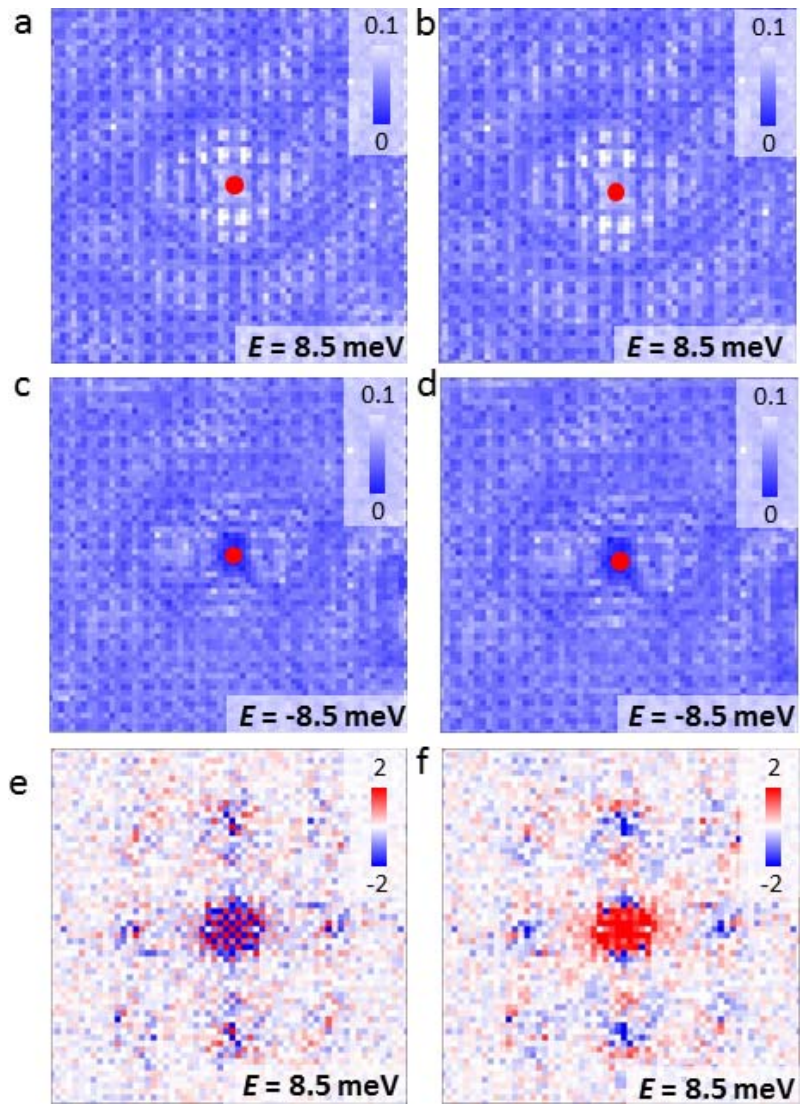

**Figure S2 | Algorithm for extracting the antisymmetric part of the FT-QPI. a, c,** Measured $g(\mathbf{r},E)$ around a single impurity for $E$=8.5, -8.5 meV. **b, d,** Edge pixels removed

to ensure that the impurity is located exactly at the center of the image. **e,** Antisymmetrized real-part is calculated as $\delta\rho^-(\mathbf{q}, E = 8.5 \text{ meV}) = \text{Re}[\rho(\mathbf{q}, E = +8.5 \text{ meV}) - \rho(\mathbf{q}, E = -8.5 \text{ meV})]$. The difference between **e** (no edge points removed) and **f** illustrates the importance of precise selection of the FT origin.

### 3. Masking out the effect of a bound state on the integral of $\delta\rho^-$.

As shown in Fig. 3**b**, the impurity bound state induces a giant peak around 4 meV in the integral of $\delta\rho^-(\mathbf{q}, E)$, which steals the weight from between the two superconducting gaps $\Delta_1 \approx 14$ meV and $\Delta_2 \approx 8.5$ meV. However, according to the theoretical proposal[S3], the behavior of $\delta\rho^-(\mathbf{q}, E)$ between the large gap and the small gap is of vital importance for distinguishing between the $s^\pm$ and $s^{++}$ pairing. To get a clear view of the signal between two superconducting gaps, we remove the central part of the d*I*/d*V* map within a circle of the radius *R*=3 pixels, which removes the states mostly affected by the impurity; instead, we use inside this circle a parabolic extrapolation of the form $\delta\rho(\omega) = A\omega^2 + B\omega$, where *A* and *B* were chosen so as to ensure continuity and smoothness of the resulting d*I*/d*V* map. After that, we calculated the integral of $\delta\rho^-(\mathbf{q}, E)$ using the resulting d*I*/d*V* map, thus obtaining the data in Fig. 3**c**.

### 4. Theoretical calculations including masking out the bound state peak

To facilitate the comparison with the experimental STM data we employed the multiband T-matrix approach using two elliptical electron bands with the dispersion

$\varepsilon_\nu(\mathbf{k}) = t\left(\frac{k_x^2}{1\pm\epsilon} + \frac{k_y^2}{1\mp\epsilon}\right) - \mu$ with $t$ = 34 meV, $\epsilon = 0.2$ and $\mu = 10.6$ meV. The scattering of quasiparticles by non-magnetic impurities, measured in the FT-QPI, was calculated as a correction to the local density of states (LDOS), using the standard T-matrix approach describing multiple scattering by a single impurity. In particular, we compute the correction to the local density of states

$$\delta\rho(\omega) = \frac{1}{2}\text{Tr Im} \sum_{\mathbf{k,q},\mu,\nu}(\tau_0 + \tau_3)G_\mu^0(k,\omega)\, t_{\mu\nu}(\omega)G_\nu^0(k,\omega) \qquad (S1)$$

Here, $G_\mu^0(k,\omega) = -\frac{i\omega\tau_0 + \varepsilon_\mu(\mathbf{k})\tau_3 + \Delta_\mu\tau_1}{\omega^2 + \Delta_\mu^2 + \varepsilon_\mu^2(\mathbf{k})}$ refers to the Nambu-Gor'kov Green's function for the band $\mu$, $\tau_i$ is the *i*-th Pauli matrix. As in the experiment, we choose the superconducting gaps to be $\Delta_{small}$ =8.5 meV and $\Delta_{large}$ =14.5 meV. The T-matrix in the band and Nambu-Gor'kov space is defined as

$$\hat{t}(\omega) = \left[1 - \hat{U}\tau_3 \sum_\mathbf{k} \hat{G}(\mathbf{k},\omega)\right]^{-1} \tau_3 \hat{U}, \qquad (S2)$$

where $U_{\mu\mu} = U_{intra}\tau_0$ and $U_{\mu\nu} = U_{inter}\tau_0$ are the intra- and interband non-magnetic impurity scattering strength, respectively. To be specific, we used the s$^\pm$ scenario, and in order to obtain a bound state we used $U_{intra} = U_{inter} = 6.8\, t$. Furthermore, for correction to the local density of states for the s$^{++}$ (red dashed curve) and s$^{\pm}$ (blue solid thick curve) scenarios. The thin blue curve corresponds to the s$^{\pm}$ case with parabolic masking of the resonant bound state in the real space as $\delta\rho(\omega) = A\omega^2 + B\omega$ with $A$ = -0.213x10$^{-3}$ states/meV$^3$ and $B$ = 2.354x10$^{-3}$ states/meV$^2$ as described in the text. **b**, The normal state correction to the local density of states for $\omega = 0$.

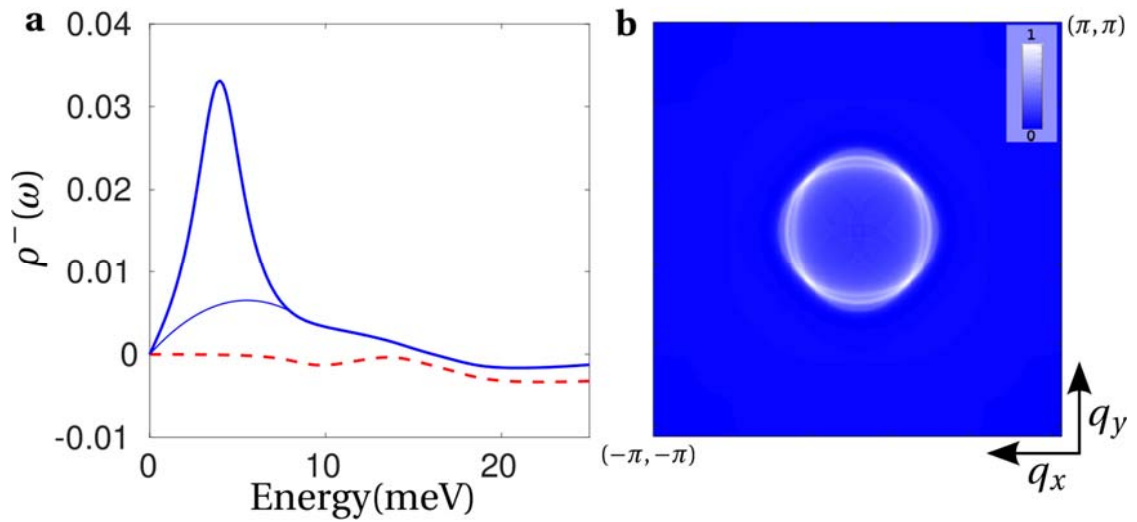

**Figure S3 | Theoretical modelling of the STM data**. **a**, Anti-symmetrized **k**-integrated

numerical stability we also assumed a quasiparticle damping of $\Gamma$=0.2$\Delta_{small}$. The results for the normal state $\delta\rho(\omega = 0)$ are shown in Fig. S3**b**.

As in the experiment, we next compute the anti-symmetrized correction to the local density of states $\delta\rho^-(\omega) = \frac{1}{2}\big(\delta\rho(\omega) - \delta\rho(-\omega)\big)$ in the superconducting state, following the original proposal of Ref. [S3] using the same (s$^{++}$) and the opposite (s$^{\pm}$) phases of the superconducting orders. The results are shown in Figure S3**a** for the energy up to 20 meV. Clearly the phase structure is strongly visible in the energy interval $\Delta_{small} < \omega < \Delta_{large}$. To exclude the effect of the impurity bound state, which occurs for the s$^{\pm}$ case at $U_{intra}$=$U_{inter}$, we masked the bound state in the real space by using in its stead a parabolic extrapolation, as shown by thin blue curve, so that the resulting function is continuous and smooth. A similar procedure was applied to the experimental data, as described in the next section.

**References**

S1. Li, W. *et al*. Phase separation and magnetic order in K-doped iron selenide superconductor. *Nature Phys.* **8**, 126-130 (2012).

S2. Sprau, P. O. *et al*. Discovery of Orbital-Selective Cooper Pairing in FeSe. Preprint at

https://arxiv.org/abs/1611.02134 (2016).

S3. Hirschfeld, P. J., Altenfeld, D., Eremin, I. & Mazin, I. I. Robust determination of superconducting gap sign changes via quasiparticle interference. *Phys. Rev. B* **92**, 184513 (2015).